\documentclass{jaa}

\usepackage{graphicx}
\usepackage{lscape}
\usepackage{deluxetable}

\begin{document}\sloppy

\title{INVESTIGATION OF THE HELIUM ENHANCEMENT IN A SUPER LITHIUM RICH GIANT HD\,77361}

\author{B. P. Hema\textsuperscript{1,*} \and Gajendra Pandey\textsuperscript{1}}
\affilOne{\textsuperscript{1}Indian Institute of Astrophysics, Koramangala II Block, Bengaluru, Karnataka-560034, India.\\}


\twocolumn[{

\maketitle

\corres{hemabp.phy@gmail.com}


\begin{abstract}

In this work, the helium-enhancement (He-enhancement) 
in the lithium-rich (Li-rich) K-giant HD\,77361 is investigated  
using the  strengths of the MgH band and the Mg\,{\sc i} lines. 
The detailed abundance analysis and also the synthesis of the MgH band
and the Mg\,{\sc i} lines has been carried out for HD 77361.
One would expect, within uncertainities, same Mg abundance from both 
the MgH and Mg\,{\sc i} lines. But, we found that Mg abundance derived 
from MgH lines is  significantly less than  the abundance 
from  Mg\,{\sc i} lines, and this difference cannot be
reconciled by changing the stellar parameters within the uncertainties,
implying He enhancement in star's photosphere.
The He enhancement in the atmospheres is 
estimated by using models of different He/H ratios so that
both the lines, MgH as well as Mg\,{\sc i}, return the same
Mg abundance for the adopted model's He/H ratio.  We found  
He/H=0.4$\pm$0.1 as the value for HD\,77361, the normal value of He/H=0.1.
Knowing the amount of He-enhancement in the Li-rich giants is 
a strong clue for understanding the scenarios responsible 
for the Li and He enrichment. The analysis and results are discussed.

\end{abstract}

}]



\doinum{12.3456/s78910-011-012-3}
\artcitid{\#\#\#\#}
\volnum{000}
\year{0000}
\pgrange{1--}
\setcounter{page}{1}
\lp{1}

\section{Introduction}

Lithium (Li) is a fragile element that gets destroyed
at temperatures of about 2.5 $\times$ 10$^{6}$ K.
The initial value of Li abundance from the interstellar medium 
is about log\,$\epsilon$(Li)=3.3 dex. Due to the mixing 
in the pre-main sequence phase, main-sequence phase,
and during the first dredge-up, the Li gets diluted and the 
 log\,$\epsilon$(Li) reduces to about 1.5-1.8 dex in the
low-mass G-K giants (Lambert {\em et al.} 1980, 
Charbonnel \& Balachandran 2000). 
This Li abundance is the upper limit 
for most of the G-K giants.  
But there is a group of G-K giants having
very strong Li lines in their spectra. 
They have unexpectedly high values of the Li abundance 
of about log\,$\epsilon$(Li) $>$ 2.5-3.0 dex.
These are the enigmatic group of stars, for which the
source/process of the Li-enhancement is not clear yet.
One of the scenarios that explains the Li-enhancement
is dredge-up from the deeper layers, where Li is
produced. 

In this work we have investigated the amount of helium (He)
in the atmosphere of one such super Li-rich giant HD\,77361.
We suspect that during the mixing process from the 
deeper layers, the convection zone dredges-up the He 
along with the other hydrogen burning elements such as 
Li, C, N, and $^{13}$C. The analysis and the results are
discussed in the following sections.

\section{Observations}

The high-resolution optical spectrum of the super Li-rich
cool giant HD\,77361 was obtained from the McDonald Observatory
by Kumar \& Reddy (2009) for their study.
The spectrum obtained with the 2.7\,m Harlan J. Smith Telescope
and Tull coude cross-dispersed echelle spectrograph
(Tull {\em et al.} 1995) is at a resolving power of about
$\lambda$/$\Delta\lambda$ = 50000. The spectra were reduced
using IRAF software package following the standard procedure.

\section{Abundance analysis}

Using the high-resolution spectra of the program star
HD 77361, the equivalent widths for several spectral
lines of several elements were measured using
tasks in the IRAF package. The equivalent widths are
measured for the weak and moderately strong, both neutral
and ionized, spectral lines.
The very strong and saturated spectral lines were discarded
as they do not fall in the linear part of the curve-of-growth
for the stellar spectral lines.The line list of 
(Johnson \& Pilachowski 2010) was used. The elements for which the
lines are very few or none in the Johnson \& Pilachowski's (2010)
list, for those the lines were adopted from 
Ram{\'{\i}}rez \& Allende Prieto (2011).

For the determination of stellar parameters and the elemental
abundances, the LTE line analysis and spectrum synthesis code
MOOG (Sneden 1973) and the ATLAS9 (Kurucz 1998) plane parallel,
line-blanketed LTE model atmospheres were used.
The microturbulence ($\xi_{t}$) is derived using Fe\,{\sc i}
lines having a similar excitation potential and a range
in equivalent width, weak to strong, giving the same abundance
(see Figure 1).
The effective temperature ($T_{\rm eff}$) is determined
using the excitation balance of Fe\,{\sc i} lines having
a range in lower excitation potential (see Figure 2). 
The $T_{\rm eff}$ and $\xi_{t}$ were fixed iteratively. 
The process was carried out until both returned zero slope. 
By adopting the determined
$T_{\rm eff}$ and $\xi_{t}$, the surface gravity (log $g$)
is derived. The surface gravity is fixed by demanding
the same abundances from the lines of different ionization
states of a species, known as ionization balance. The surface
gravity is derived using the lines of Fe\,{\sc i}/Fe\,{\sc ii},
Ti\,{\sc i}/Ti\,{\sc ii}, and Sc\,{\sc i}/Sc\,{\sc ii}.
Then, the mean log $g$ was adopted (see Figure 3).

\begin{figure}
\includegraphics[width=1.02\columnwidth]{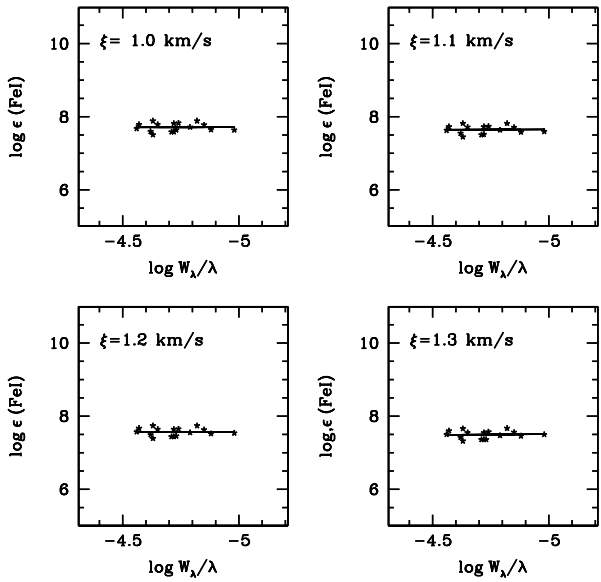}
\caption{Estimation of the microturbulence, $\xi_{t}$ for HD\,77361.}
\label{}
\end{figure}

\begin{figure}
\includegraphics[width=1.03\columnwidth]{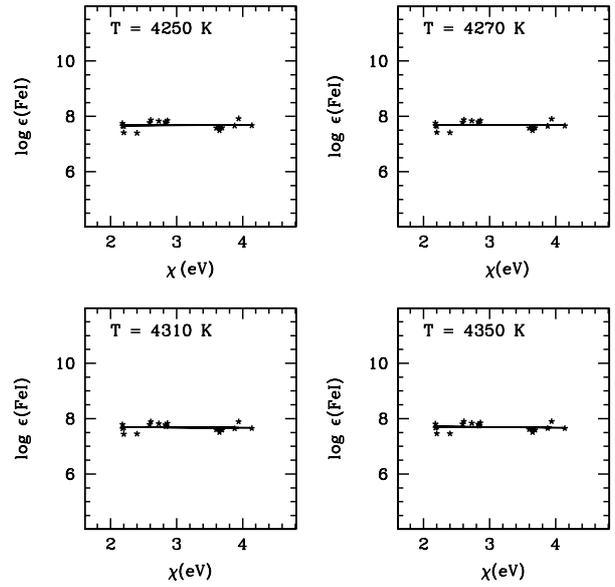}
\caption{Estimation of the effective temperature,
$T_{\rm eff}$ for HD\,77361.}
\label{}
\end{figure}

\begin{figure}
\includegraphics[width=1.02\columnwidth]{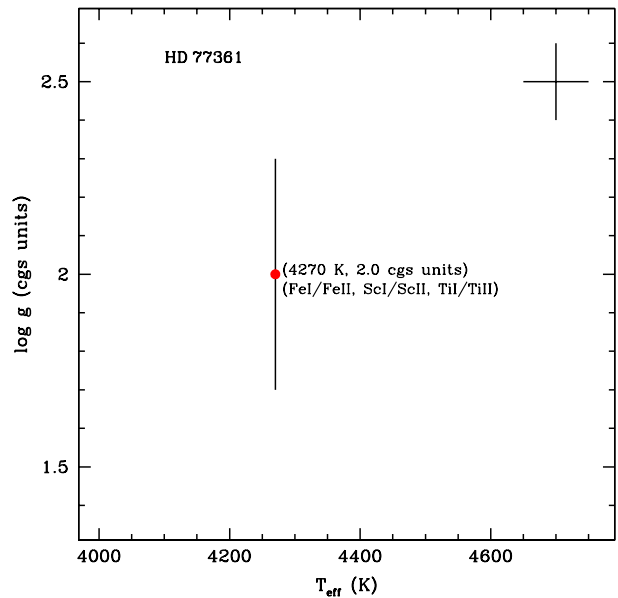}
\caption{Estimation of the surface gravity,
log $g$ for HD\,77361.}
\label{}
\end{figure}

For HD\,77361, the stellar parameters are determined by
Kumar \& Reddy (2009) and also by Lyubimkov {\em et al.} (2015). 
The stellar parameters determined spectroscopically 
by Kumar \& Reddy (2009)
and Lyubimkov {\em et al.} (2015), respectively, are;
($T_{\rm eff}$, $\log g$, $\xi_{t}$, log $\epsilon$(Fe)):
(4580$\pm$75\,K, 2.5$\pm$0.1 cm\,s$^{-2}$, 1.4$\pm$0.5 km\,s$^{-1}$,
7.48$\pm$0.1), and  (4370$\pm$100\,K, 2.30$\pm$0.1 cm\,s$^{-2}$,
1.1$\pm$0.2 km\,s$^{-1}$, 7.49$\pm$0.14).
The stellar parameters derived in this work are:
$T_{\rm eff}$ = 4270$\pm$50\,K, $\log g$ = 2.0$\pm$0.1 cm\,s$^{-2}$,
$\xi_{t}$ = 1.2$\pm0.1$ km\,s$^{-1}$, [Fe/H]=7.45$\pm$0.12. 
The stellar parameters are in very good  agreement
with those derived by Lyubimkov {\em et al.} (2015).
Adopting these stellar parameters, the MgH bands and
the neutral Mg lines are analysed. 

Uncertainties on the $T_{\rm eff}$ and $\xi_{t}$ are estimated 
by changing the $T_{\rm eff}$ in steps of 25 K and $\xi_{t}$ 
in steps of 0.05 km\,s$^{-1}$. The change in $T_{\rm eff}$ 
and $\xi_{t}$ and the corresponding deviations in abundance,
from the zero slope abundance, of about 1$\sigma$ error, is obtained.
This change is adopted as the uncertainty on these parameters.
The adopted $\Delta$$T_{\rm eff}$ = $\pm$50\,K and
$\Delta$$\xi_{t}$ = $\pm$0.1\,km\,s$^{-1}$ (see Figure 4).
The uncertainties on log $g$ is the standard deviation
from the mean value of the log $g$ determined from
different species, which is about $\pm$0.1 (cgs units).
The linelist used for the abundance analysis 
is adopted from Hema {\em et al.} (2018). 
 
The elemental abundances derived for the determined stellar 
parameters for the normal He/H ratio 0.1, is given in Table 1. 
The solar abundances (Asplund {\em et al.} 2009),  
the abundance ratios ([E/Fe]) and the number of lines used
are also given. 
An Mg\,{\sc i} line at 5711\AA\ in HD\.77361 is also
synthesized to support the Mg abundance derived from Mg\,{\sc i}
equivalent width analysis (see Figure 5).

\begin{figure}
\includegraphics[width=1\columnwidth]{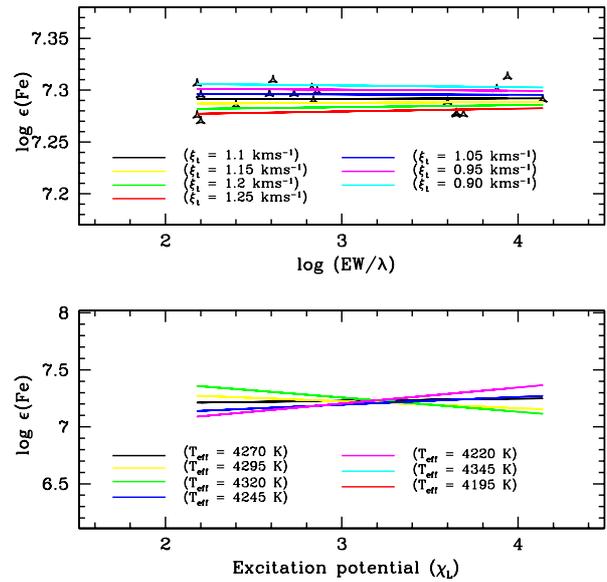}
\caption{Error analysis on the effective temperature ($T_{\rm eff}$)
and the microturbulence ($\xi_{t}$) for HD\,77361.}
\label{}
\end{figure}

\begin{figure}
\includegraphics[width=0.9\columnwidth]{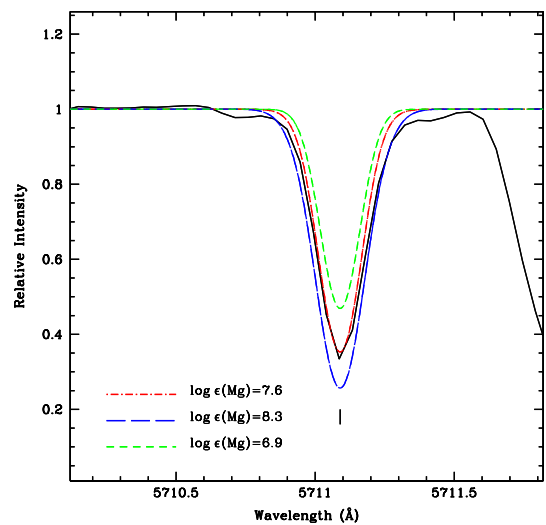}
\caption{The synthesis of Mg\,{\sc i} line
at 5711\AA\ for HD\,77361.  The synthesis is
shown for the best fit Mg abundance
and also for other two Mg abundances for comparison.}
\label{}
\end{figure}

\section{MgH Band and Spectrum Synthesis}

Using the derived stellar parameters and the
elemental abundances, the MgH band in
the spectra of HD\,77361
is analysed. Our aim is to investigate if
the star is He-enhanced. Similar to our previous
studies, using the procedure that was developed by
Hema \& Pandey (2014) and Hema {\em et al.} (2018)
the He-enhancement is investigated.
Our method is that, the Mg abundance derived
from the subordinate lines of (0,0) MgH molecular band 
that are free from contamination, 
and that from the Mg\,{\sc i} atomic lines derived from the  
neutral weak and clean Mg\,{\sc i} lines, are compared to examine 
if there exists any difference between these two abundances
beyond the uncertainties.  

The Mg abundance from the atomic lines are derived from
the equivalent width analysis and also by 
synthesis. The well-identified Mg\,{\sc i} line
at 5711\AA\ is synthesized and is shown in Figure 5. 
The derived Mg abundance by equivalent width analysis 
and by synthesis are; log $\epsilon$(Mg) = 7.45$\pm$0.13 
and 7.6 dex, respectively. 
These Mg abundances are in good agreement within the 
uncertainties.

Spectrum of the (0,\,0) MgH band is synthesized
from 5100--5200\AA\ which includes the Mg\,b lines
and the (0,\,0) MgH band. The procedure adopted here
is from Hema \& Pandey (2014) and Hema {\em et al.} 2018. 
For synthesizing
the spectra, the atomic lines were compiled from the
standard atomic data sources, and all of the atomic
lines identified by Hinkle {\em et al.} (2000) were included. 
The (0,\,0) MgH molecular line list was adopted from 
Hinkle {\em et al.} (2013). Synthetic
spectra were generated by combining the LTE spectral
line analysis/synthesis code MOOG (Sneden 1973), and the
ATLAS9 (Kurucz 1998) plane parallel, line-blanketed LTE
model atmospheres with convective overshoot. Spectrum of
Arcturus, a typical red giant, was synthesized to validate the
adopted gf--values of the atomic/molecular lines. Using the
spectrum synthesis code, synth in MOOG, the high-resolution
optical spectrum of Arcturus was synthesized for the
stellar parameters and the abundances given by 
Ram{\'{\i}}rez \& Allende Prieto (2011).
The synthesized spectrum was convolved with a Gaussian
profile with a width that represents the broadening due to
macroturbulence and the instrumental profile. Minimal adjustments
were made to the abundances of the atomic lines to obtain
the best fit to the observed high-resolution optical spectrum of
Arcturus (Hinkle {\em et al.} 2000). The changes in the log gf--values
were not more than 0.1 dex. A reasonably good fit was obtained
to the MgH molecular lines for the adopted isotopic values from
McWilliam \& Lambert (1988).

A good match of the synthesized Arcturus spectrum to the
observed high--resolution spectra validates the
adopted line list for the adopted stellar parameters of Arcturus.
These checks on the published analysis of Arcturus are taken
as evidence that our implementation of the code MOOG, the
LTE models, and the adopted line list were successful for
the syntheses of the red giants’ spectra. Hence, the spectrum 
of HD\,77361 was synthesized following the above procedure.

\begin{table*}[htb]
\begin{center}
\caption{Abundances for Different He/H Ratios for HD\,77361.}
\begin{tabular}{lccrcrr}
\tableline\tableline
Elements & log $\epsilon\odot$ & \multicolumn{5}{c}{HD\,77361}\\
&   & log $\epsilon$(He/H=0.1)  & [X/Fe] & log $\epsilon$(He/H=0.40)&
[X/Fe]   & $n$\\
\tableline
H & 12.00 & 12.00 & \nodata & 11.735 & \nodata  &\nodata \\
He & 10.93 & 11.00 &  \nodata & 11.337 & \nodata  &\nodata\\
Li & 1.05 & 3.90 & 2.9 & 3.67 & 3.02 & 1\\
C & 6.24 & 8.7$\pm$0.14 & 2.51 & 7.74$\pm$0.15 & 1.90 & 3\\
O & 8.69 & 8.66$\pm$0.02 & 0.02 & 8.45$\pm$0.03 & 0.16 & 2 \\
Na & 6.24 & 6.18$\pm$0.05 & $-$0.01 & 5.89$\pm$0.05 & 0.05 & 3\\
Mg (Mg\,{\sc i}) & 7.60 & 7.45$\pm$0.13 & $-$0.10 & 7.05$\pm$0.09 & $-$0.15 & 4 \\
Mg (MgH) & \nodata &  7.00 & $-$0.55 & 7.05 & $-$0.15 & \nodata\\
Al & 6.45 & 6.40$\pm$0.08 & 0.00 & 6.05$\pm$0.08 & 0.00 & 4 \\
Si & 7.51 & 7.80$\pm$0.12 & 0.34 & 7.30$\pm$0.11  & 0.19 & 5\\
Ca & 6.34 & 5.93$\pm$0.18 & $-$0.36 & 5.58$\pm$0.17 & $-$0.36 & 8\\
Sc\,{\sc i} & 3.15 & 2.81$\pm$0.13 & $-$0.29 & 2.69$\pm$0.12 & $-$0.06 & 3\\
Sc\,{\sc ii} & 3.15 & 3.00$\pm$0.12 & $-$0.10 & 2.69$\pm$0.07 & $-$0.06 & 5 \\
Ti\,{\sc i} & 4.95 & 4.48$\pm$0.05 & $-$0.42 & 4.31$\pm$0.06 & $-$0.24 & 6\\
Ti\,{\sc ii} & 4.95 & 4.62$\pm$0.08 & $-$0.28 & 4.24$\pm$0.10 & $-$0.31 & 4 \\
V & 3.93  & 3.56$\pm$0.10 & $-$0.32 & 3.34$\pm$0.10 & $-$0.19 & 4\\
Cr & 5.64 & 5.32$\pm$0.14 & $-$0.27 & 5.03$\pm$0.12  & $-$0.21 & 7 \\
Mn & 5.43 & 5.30$\pm$0.20 & $-$0.08 & 5.06$\pm$0.19 & $-$0.03 & 3\\
Fe\,{\sc i} & 7.50 & 7.45$\pm$0.12 & $-$0.05 & 7.10$\pm$0.12 & $-$0.40 & 18\\
Fe\,{\sc ii} & \nodata & 7.50$\pm$0.05 & 0.00 & 7.00$\pm$0.05 & $-$0.50 & 4\\
Co & 4.99 & 5.13$\pm$0.17 & 0.19 & 4.78$\pm$0.17 & 0.19 & 6 \\
Ni & 6.22 & 6.19$\pm$0.18 & 0.02 & 5.81$\pm$0.13 & $-$0.01 & 3 \\
La & 1.10 & 2.10 & 1.05 &  1.76 & 1.06 & 1\\
\tableline
\end{tabular}
\tablenotes{{\bf Note.} Abundances derived for the normal (He/H=0.1) and
the determined (He/H) for the program star is given. The
corresponding abundance ratios, the number of spectral lines
used, and the solar abundances (Asplund et al. 2009) are
also provided.}
\end{center}
\end{table*}

\begin{figure*}[!htb]
\includegraphics[scale=0.80]{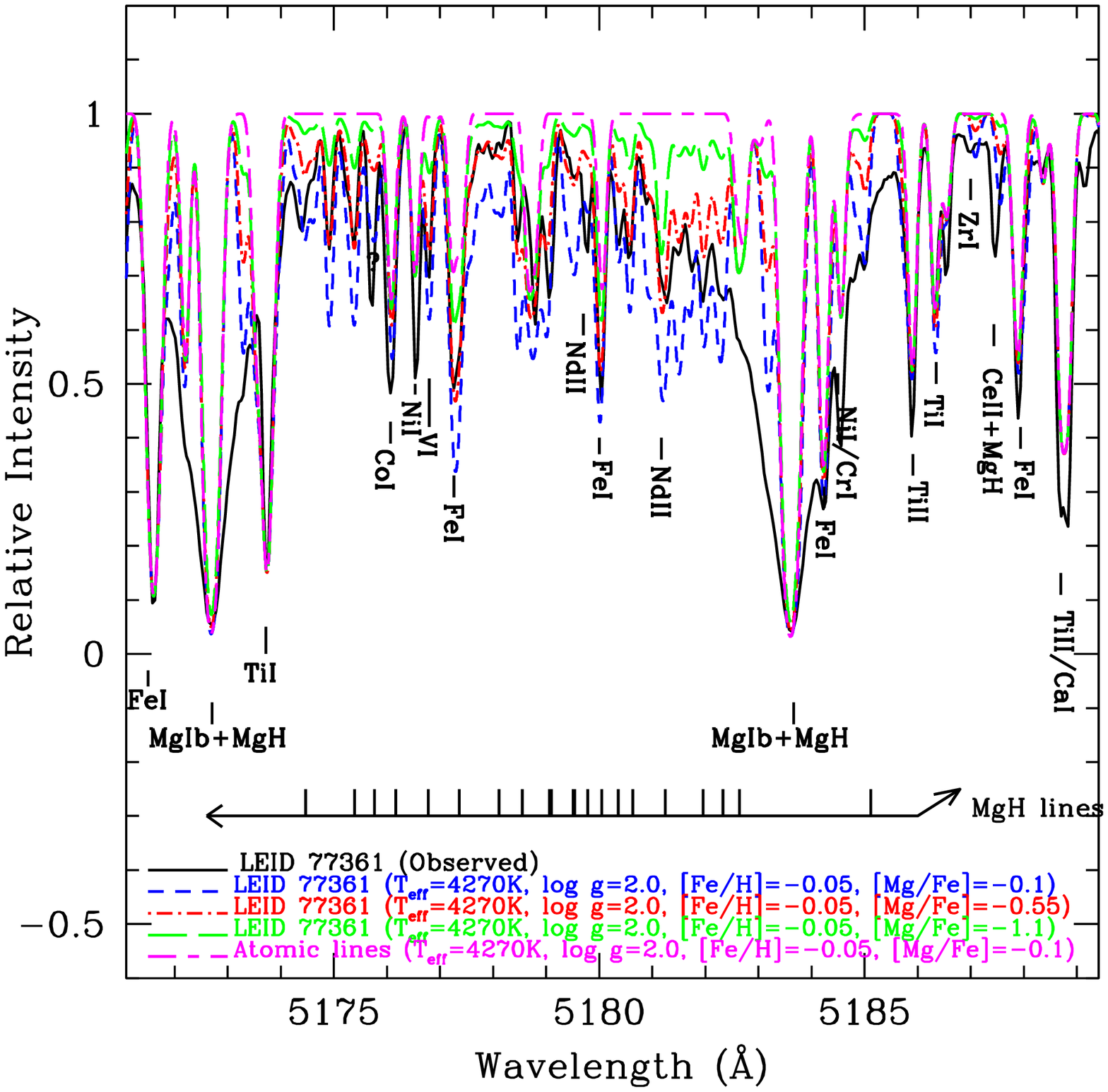}
\caption{Superposition of the observed and the synthesized
spectra for the program star HD 77361. The spectrum
is synthesized for the star’s derived stellar parameters
and the Mg abundance. The synthesis is shown for
the best-fit [Mg/Fe] = $-$0.55 with the red dot-dashed line,
the synthesis for [Mg\,{\sc i}/Fe] = $-$0.1 is shown with the blue
short-dashed line, and the synthesis for [Mg/Fe] = $-$1.1
is shown with the green long-dashed line, for comparison.
The synthesis for pure atomic lines is also shown with the
magenta short/long-dashed line. The key lines are marked.}
\label{}
\end{figure*}

The Mg abundance derived from the subordinate lines of
MgH band for the stellar parameters derived (see Section 3)
is about 7.0, while the Mg abundance derived from the
Mg\,{\sc i} lines is about 7.45$\pm$0.1 dex (see Figure 6).
The abundance ratio [Mg/Fe] derived from MgH bands is less 
by about $-$0.45 dex than that derived from the Mg\,{\sc i} lines. 
The Mg abundance from the subordinate lines of the MgH band
were derived by changing the stellar parameters 
within the uncertainties and is summarized in Table 2. 
From Table 2, it is clear that this difference cannot be reconciled 
by making the changes to the stellar parameters within the 
uncertainties (see Figure 6) and also for slightly higher margin
on the uncertainty of the $T_{\rm eff}$. 
The Mg abundance derived from the MgH bands with a change of about 
$\pm$100 K on the derived $T_{\rm eff}$ also do not provide a 
comparable Mg abundance from that of the Mg\,{\sc i} lines.
Hence, this difference is attributed to the lower hydrogen or 
He-enhancement in the atmosphere of the program star. 
The amount of He-enhancement is determined using the
appropriate models and is discussed in the following section.

\begin{table*}[!htb]
\begin{center}
\caption{The derived Mg abundances from Mg\,{\sc i} lines and the
MgH band for the adopted stellar parameters (in boldface) and 
the uncertainties on them. }
\begin{tabular}{cccccccccccc}
\tableline\tableline
$T_{\rm eff}$ &  log $g$ & [Mg/Fe] from Mg\,{\sc i} & [Mg/Fe] from MgH 
\\
\tableline
{\bf 4270} & {\bf 2.0} & {\bf $-$0.1} & {\bf $-$0.55} \\
4320 & 2.0 & $-$0.13 & $-$0.50 \\
4220 & 2.0 & $-$0.11 & $-$0.70 \\
4270 & 2.1 & $-$0.09 & $-$0.60 \\
4270 & 1.9 & $-$0.12 & $-$0.50 \\
4170 & 2.0 & $-$0.09 & $-$0.70 \\
4370 & 2.0 & $-$0.13 & $-$0.30 \\
\tableline
\end{tabular}
\end{center}
\end{table*}

\begin{figure*}[!htb]
\includegraphics[scale=0.80]{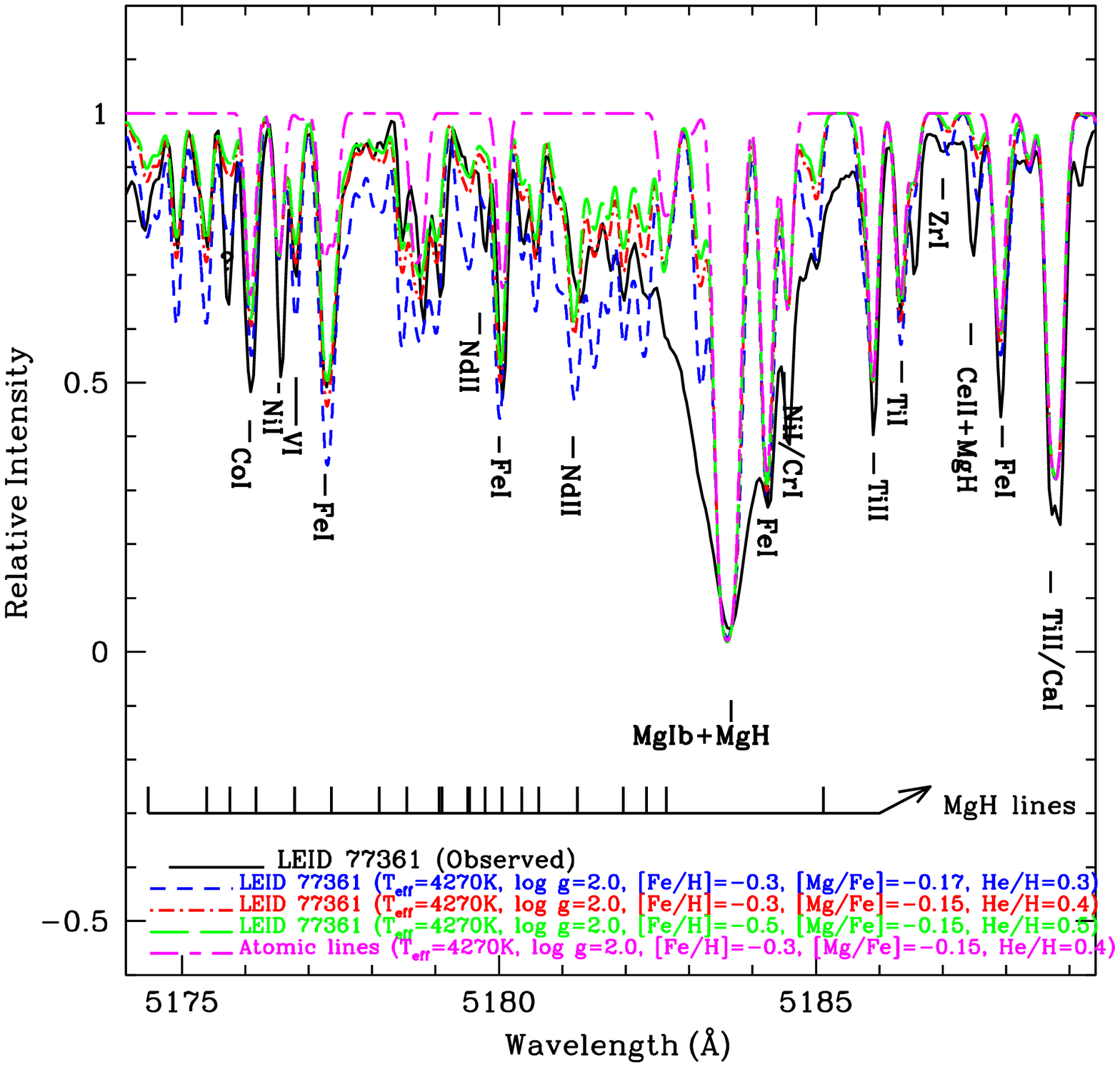}
\caption{Observed and the synthesized MgH bands for 
HD\,77361 are shown. The spectra synthesized for the 
Mg abundance derived from the Mg I lines and the 
best-fit value of He/H ratio are shown by the red 
dash$-$dotted line. The synthesis for the two value 
of the He/H are also shown.}
\label{}
\end{figure*}

\section{Determination of the He/H ratio for HD\,77361}

For the abundance analysis discussed in section 3,
the input abundances of H and He provided to MOOG, 
that adopts a model atmosphere computed for a normal 
He/H ratio of 0.1, are log $\epsilon$ (H)=12 and
log $\epsilon$ (He)=11, respectively. 
The observed spectra of the program star was reanalysed
using model atmospheres with differing He/H ratios, 
computed and interpolated for different He/H ratios 
(Kurucz 1993, Kurucz 2014, Kurucz 2017, Kurucz 2018
Sbordone {\em et al.} 2004, 2007, Sbordone 2005,
Castelli \& Kurucz 2003), For details on model 
atmosphere see section 3.2 of Hema {\em et al.} (2020).

The stellar parameters were rederived using a grid of 
model atmosphere with He/H=0.20, 0.30, 0.40, 0.50;
the adopted metallicity of the grid is fixed based on the derived
metallicity of the program star for the normal He/H ratio.
The procedure adopted for determining the stellar parameters 
is ditto as described in Section 3. The rederived stellar parameters
are not sensitive and almost independent of the
adopted grid's He/H ratio.  
For the program star's derived stellar parameters, synthetic spectra
in the MgH band region were computed for different He/H ratios.
The Mg abundance used for the above
synthesis were derived from the measured equivalent
widths of the weak Mg\,{\sc i}
lines for the adopted model's He/H ratio. Hence, the best fit
to the MgH band in the observed spectrum determines the
adopted model's He/H ratio.
Finally, the elemental abundances are derived for
the adopted stellar parameters: $T_{\rm eff}$,
$\log g$, $\xi_{t}$, and the He/H ratio (see Table 1). Note that
the adopted metallicity of the model atmosphere comes from
the iron abundances derived from the measured equivalent widths of
the Fe lines for the program star, and is an iterative process.

In the case of elemental abundances derived
by adopting model atmospheres
with same stellar parameters but varying He/H ratios (for example,
He/H=0.20 and 0.10), the abundances from the higher
He/H ratios are lower than that derived from the normal.
Here, decreasing the hydrogen abundance or increasing the
helium abundance i.e., increasing the He/H ratio,
lowers the continuous opacity per gram (Sumangala Rao {\em et al.} 2011). 
Hence, for the same observed strength of the spectral line,
the elemental abundance has to decrease (see Table 1).
The decrease in abundance is proportional to the amount
of H-deficiency or the He-enrichment applied in the sense of
He/H ratio.
However, the abundance ratios remain unchanged for most
of the elements with few exceptions.

Examples of the spectrum syntheses in the MgH band region for
the program star are presented in figure 7.
By adopting the respective elemental abundances derived from
the adopted model computed for the pair of [Fe/H] and the He/H ratio,
spectra in the MgH band region were synthesized.
Figure 7, for HD 77361, clearly shows that
for the Mg abundance of 7.05$\pm$0.09 (from Mg\,{\sc i} lines), the
best fit to the observed MgH band (mainly the subordinate
lines of MgH band at about 5175\AA) is obtained for He/H
ratio, He/H=0.40 and the metallicity, [Fe/H]=$-$0.3.
Nevertheless, the input abundances
of H and He provided to MOOG are log $\epsilon$ (H)=11.735 and
log $\epsilon$ (He)=11.337, respectively, for the adopted 
model of He/H=0.4,
obtained using the standard
relation: 
$n_{\rm H}$+ 4 $n_{\rm He}$=10$^{12.15}$.
And, the corresponding mass fractions for H and He
are: Z(H)=0.385 and Z(He)=Y=0.614.

The He/H ratio for the program star is determined from
the observed MgH band in their spectra. Once the stellar
parameters and metallicity are determined and fixed,
the MgH band strength depends mainly on the
atmosphere's Mg abundance and the He/H ratio.
Then, the uncertainty on the He/H ratio is primarily due to
the uncertainty on the Mg abundance, which is about 0.1.
Hence,  for the program star HD\,77361, He/H\,=\,0.40$\pm$0.1. 

\section{Discussion}

HD\,77361 is a  super Li-rich K-giant having high value of
observed Li-abundance like the known Li-rich giants. This 
is located around the bump region of the red giant branch.
Our determination of Li-abundance which is about 3.9
is obtained by a single line that is 6103\AA, since the 
Li\,{\sc i} line 6707\AA\ is very strong. 
This is in excellent agreement with that derived by 
Kumar \& Reddy (2009). Our aim was to investigate the 
He-enhancement in the Li-rich giants. Comparing the abundances
derived from the Mg\,{\sc i} lines and from the subordinate lines 
of the (0,0) MgH band, the He abundance (He/H ratio) is estimated 
for HD\,77361. Using the appropriate models, the He-enhancement 
is determined. The He-enhancement is about He/H=0.4$\pm$0.1,  
the normal value (He/H=0.1). 
By adopting the derived He-abundance(He/H ratio) and the 
metallicity [Fe/H], 0.4$\pm$0.1 and $-$0.3, respectively,
the corresponding elemental abundances are derived. Using 
the He-enhanced models, the Li abundance is about 3.7, which 
still places HD\,77361 among the super Li-rich giants.  

The two plausible scenarios for Li-enrichment in G- and K-giants 
are, the dredge-up from deeper layers which surfaces the 
hydrogen-burning products, and the planet engulfment.
Since, HD\,77361 is a super Li-rich giant and is located around
the RGB-bump, the potential scenario for Li-enrichment 
is the dredge-up process. Since the convective zone 
reaches into the deeper layers, the mixing dredges up the 
hydrogen-burning products  such as Li, C, N, and also 
$^{13}C$ to the surface.
From our analysis, HD\,77361 is also enhanced in helium 
and is a product of hydrogen burning. Hence, it appears
that along with the Li, and other hydrogen burning products, 
He is also brought up to the surface due to extra mixing 
by deeper convective zone. This could be one of the possible 
scenarios for Li- and He-enrichment in HD\,77361.
However, the He-enhancement in the Li-rich giants located at 
different positions in the RGB-phase, such as RGB-bump, clump, 
RGB-ascent and also at the RGB-tip, needs to be explored,
and this work is in progress. 
Knowing the He- and Li-enrichment in these stars 
would be a strong clue for understanding the 
processes responsible for these enhancements.

We are very thankful to  Dr. Bharat Kumar Yerra 
and Prof. Eshwar Reddy for providing us with the spectra 
of HD\,77361. We thank Prof. Carlos Allende Prieto for 
helping us in obtaining the He-enhanced models.


\end{document}